# But, how can the atom be so stable, Dr. Maxwell?

Costas J. Papachristou

Department of Physical Sciences, Hellenic Naval Academy

papachristou@hna.gr

***Abstract.*** In the initial stages of its development, atomic theory had to bypass the laws of classical electromagnetism in an *ad hoc* manner in order to explain the stability of atoms. In quantum mechanics, however, the classical theory may find again some room even for a microscopic structure such as the atom. Provided, of course, that certain classical concepts are reexamined and suitably reinterpreted...

## *1. Electromagnetic radiation: a triumph of classical Physics*

There is no doubt that *James Clerk Maxwell* (1831-1879) was the leading figure of Theoretical Physics in the nineteenth century. Among his many achievements, Maxwell unified electricity and magnetism into a single electromagnetic theory and predicted the existence of electromagnetic waves. Unfortunately, Maxwell didn't live long enough to see the experimental confirmation of his prediction...

We often tend to think of electricity and magnetism as separate natural phenomena. And, indeed, they exhibit fundamental differences. For example, even stationary electric charges "feel" the electric interaction whereas only moving charges are subject to the magnetic interaction. In a hypothetical world where all electric and magnetic fields were static (i.e., time-independent) there would be no way of knowing that electric and magnetic phenomena are interrelated and mutually dependent. From a mathematical point of view, the famous four Maxwell's equations would split into two independent pairs corresponding to the electric and, separately, the magnetic field (see, e.g., Chap. 9 of [1]).

In 1831, however, Michael Faraday experimentally discovered something interesting: a time-change of a magnetic field is necessarily accompanied by the appearance of an electric field! Despite the lack of experimental evidence at his time, Maxwell predicted that the converse was also true; namely, a magnetic field appears each time an electric field changes with time. No absolute separation is thus possible between electric and magnetic phenomena, given that the electric and the magnetic field appear to be intimately related.

Historically speaking, this has been the first unification theory of seemingly independent interactions – the electric and the magnetic – into a single *electromagnetic* (e/m) interaction. In the twentieth century there would be a further enhancement of the unification scheme with the inclusion of the weak and the strong interaction, along with a heroic effort of incorporating gravity as well.

With his mathematical genius, Maxwell was able to describe the electromagnetic phenomena in terms of a set of equations that bear his name. The four *Maxwell's equations* [1] describe the behavior of the electromagnetic field in space and time. From these equations there follows the interesting conclusion that the electromagnetic field has wavelike properties. That is, a change (or, as we say, a *disturbance*) of the field at some point of space is not felt instantaneously at other points but propagates in the form of an *electromagnetic wave* (or e/m wave, for short) traveling at the speed of light. Light itself is a special kind of e/m wave having the property that it may be sensed by our eyes.

The propagation of energy by means of e/m waves is called *electromagnetic (e/m) radiation*. A physical system that emits energy in the form of e/m waves is said to *emit e/m radiation* (or, simply, to *radiate*). Examples of radiating systems are atoms, molecules, nuclei, hot bodies, antennas of radio and TV stations, etc.

By the Maxwell equations it follows that, in principle, the e/m radiation is produced in either of two ways: by *accelerated electric charges* (regarded as isolated quantities) or by *time-varying electric currents*. In particular, a charge moving at constant velocity (i.e., executing uniform rectilinear motion) does *not* radiate. In a previous article [2] we explained this by using a parable:

> On a hot summer day you go to the store and buy an ice cream. You decide to eat it on the road before it melts. You take a carefree walk on a straight path, with steady step (thus, with *constant velocity*), without noticing a swarm of bees following you (or, rather, your ice cream)! When you suddenly notice them, you *accelerate* your motion in order to escape from them (you either move faster in the same direction or just change your direction of motion). Scared by this move of yours, some of the bees leave the swarm and fly away, never to come back...

> What is the meaning of all this? The "ice cream" is an electric charge initially moving with constant velocity and carrying with it the total energy of its e/m field (the "swarm of bees"). This is just a transfer of a constant amount of energy in the direction of motion of the charge. When the charge accelerates, a part of this energy (the "bees" that fly away) is detached, in a sense, and travels to infinity at the speed of light in the form of an e/m wave. And, the higher the acceleration of the charge, the greater the energy radiated per unit time.

One might now ask the following question: As everyone knows, acceleration is always defined relative to some observer. If a charge accelerates relative to a "stationary" observer, this observer will see the charge emitting e/m radiation. However, relative to an observer moving with the charge, this charge is stationary (thus non-accelerating). How should the moving observer interpret the emitted radiation?

At this point we must recall the notion of an *inertial frame of reference* [3]. This is a system of coordinates (or axes) relative to which a free particle – i.e., a particle subject to no forces – either moves with constant velocity (executes uniform rectilinear motion) or otherwise is at rest. An observer using such a frame of reference is said to be an *inertial observer*. In accordance with the law of inertia (Newton's first law) an inertial observer moves with constant velocity (does not accelerate) relative to any other inertial observer.

What makes inertial frames really special is the fact that it is only in such frames that Newton's laws, as well as the laws of electromagnetism, are valid. In particular, an electric charge emits e/m radiation only when it accelerates relative to an *inertial* observer. An observer moving with this charge, however, is *not* inertial. Therefore, although relative to that observer the charge seems to be at rest (hence non-accelerating) the observer must still not attempt to interpret electromagnetic phenomena according to the Maxwell equations, since this would lead to the erroneous conclusion that even a charge at rest may emit radiation! In reality, of course, the charge radiates because it accelerates *with respect to the inertial observer*.

It is interesting that special relativity provides a simple proof that a charge moving with constant velocity relative to an inertial observer does not radiate. Here is this proof:

Consider a charge $q$ moving with constant velocity relative to an inertial observer $O$. Consider also an observer $O´$ who is moving with the charge. This latter observer is also inertial since she moves with constant velocity relative to $O$. Because $q$ is at rest relative to $O´$, that observer will record just a static electric field and no e/m radiation from $q$. (We remark that e/m radiation requires a *time-varying* e/m field; see, e.g., Chap. 10 of [1].)

Let us now make the assumption that the "stationary" observer $O$, relative to whom the charge $q$ moves with constant velocity, sees $q$ emitting radiation. According to the principle of relativity, e/m radiation propagates with the same speed $c$ (the speed of light) in all inertial frames of reference. Thus, if the observer $O$ records radiation propagating with speed $c$, then the observer $O´$ must also record radiation propagating with the same speed. But, as we said before, the observer $O´$ does not see any radiation whatsoever! The reason for arriving at a wrong conclusion is our initial assumption that the observer $O$ sees $q$ emitting radiation. We thus conclude that $q$ cannot emit if it moves with constant velocity with respect to the inertial observer $O$.

We note that the above line of reasoning is no longer valid if $q$ accelerates relative to $O$, since the observer $O´$ who moves with the charge is now not inertial and the principle of relativity cannot be used to correlate the observations of $O$ and $O´$.

## *2. Classical Physics and atomic theory: a problematic relationship*

An atomic system consists of a number of positively and negatively charged particles (the nucleus and the electrons, respectively) held together by electric forces in a manner that the system be stable (in the sense that it retains its identity) over a long period of time.

According to a theorem by Earnshaw, a system of charged particles cannot be in a state of stable static equilibrium under the sole action of electrostatic forces. The particles must therefore be in motion and, since this motion necessarily takes place within a very limited space, the direction of their velocity must be constantly changing. In other words, the particles must have at least a centripetal acceleration.

Now, here is the problem: According to classical electromagnetism, every accelerating charge emits e/m radiation, constantly losing energy in the process. Thus the classical theory predicts that, within a very short time interval the system must shrink and eventually collapse, losing its identity. Fortunately this never happens in reality, as the atomic systems are stable!

Another effect the classical theory is not able to explain is that the atomic systems emit and absorb e/m radiation in a selective manner. That is, each system absorbs and emits very specific frequencies of radiation. As we say, the absorption and emission spectra of the system are *line spectra*.

Where the classical theory fails, the quantum theory takes over. Let us see how this happens, taking as an example the simplest atomic system: the hydrogen atom. As a preliminary step, let us explain once more why such a system cannot be studied in the context of classical Physics.

### *3. Rutherford's model of the atom: an important beginning with incorrect conclusions*

The first modern model of the atom was proposed in 1911 by Ernest Rutherford. In the simplest case of the hydrogen atom the sole electron revolves about the nucleus (proton) in a circular orbit of arbitrary radius, having constant angular velocity.

The picture is reminiscent of the motion of a planet around the Sun, or the motion of a satellite around a planet. There is, however, a basic difference. In the case of the hydrogen atom the motion is governed by an e/m interaction (the Coulomb force between electron and proton), not by gravity. And, in view of the electron's centripetal acceleration the classical theory predicts that the atom must constantly emit e/m radiation. As a result of this loss of energy the radius of the electronic orbit must decrease continuously (cf. Chap. 1 of [1]) until finally the electron will fall into the nucleus and the atomic structure will collapse in about $10^{-8}$ seconds! This, of course, does not agree with the physical observation that the hydrogen atom is stable.

But, this is not the end of the story. During a continuous change of the size and the energy of the atom, the frequency of the emitted radiation must also change in a continuous manner [1]. As mentioned previously, however, the atoms do not emit e/m radiation within a continuous spectrum of frequencies but, instead, each atom emits a specific set of frequencies that constitutes a hallmark of the atom. In other words, the emission spectra of atoms (and likewise of molecules) are *line spectra*.

So, although an important first step toward understanding atomic structure, Rutherford's model can explain neither the stability of the atom nor the non-continuity of the atomic spectra. And here comes quantum theory – with its own initial problems...

### *4. The Bohr model: an amalgam of classical and quantum ideas*

In 1913 Niels Bohr presented a modification of the Rutherford model for the hydrogen atom by proposing a new model that combined classical concepts, such as the trajectory of a particle, with novel ideas like the quantization of angular momentum and energy.

In a rather *ad hoc* manner, Bohr enhanced the Rutherford model by adding two quantum rules:

1. The electron is not allowed to follow arbitrary circular paths around the nucleus but, instead, it must describe orbits of well-defined radii. Along these orbits the electron does *not* emit e/m radiation and the energy of the hydrogen atom assumes specific, constant values.

2. The atom emits radiation only when the electron falls from an orbit of higher energy to a smaller orbit of lesser energy. The energy is emitted in the form of a single *photon* (a quantum of e/m radiation).

Bohr's theory was able to explain the line spectrum of hydrogen, giving correct values for the observed frequencies of the emitted radiation. The line property of the spectrum can be understood in the following way: In a transition of the electron from an orbit of energy *E* to an orbit of lesser energy *E´* the atom emits a photon of frequency $v=(E–E´)/h$, where *h* is Planck's constant. And, since *E* and *E´* assume *discrete* rather than arbitrary values (that is, the energy of the atom is *quantized*), the same must be true with regard to the frequencies *v* of the emitted e/m radiation. We thus conclude that the line property of the emission (and likewise the absorption) spectrum is a direct consequence of the quantization of energy.

Bohr's model is not free from problems. Here are two major ones:

1. While it correctly explains the emission spectrum of the hydrogen atom, it cannot do the same thing for atoms having two or more electrons.

2. It does not answer the question of why the electron does not emit radiation when moving on the Bohr orbits despite its having centripetal acceleration.

Both these issues are treated successfully by quantum mechanics. It is the second, stability issue on which we will concentrate.

### *5. How quantum mechanics explains the stability of Bohr's atom*

According to classical electromagnetism, a point charge in uniform circular motion emits radiation because of its centripetal acceleration. On the contrary, a constant circular current does *not* radiate since the e/m field it produces is only a static magnetic field. As mentioned earlier, the existence of e/m radiation requires that the underlying e/m field be time-dependent (sources of static fields do not radiate).

In quantum mechanics, however, the picture of a point charge moving in a definite way on a well-defined orbit is meaningless since, by the *uncertainty principle*, it is not possible to know the exact position and velocity of an elementary particle. Instead of classical orbits, quantum mechanics speaks of *stationary states* of well-defined energies. And, the motion of an electron on a definite path around the nucleus is replaced by a *probability current* related to the possible positions the electron may occupy. When the electron is in a stationary state, the corresponding probability current is constant in time.

Moreover – and this is a crucial step – the probability current may be considered as proportional to an actual electric current around the nucleus.[1] In a stationary state this latter current is constant in time. And, classically, a constant current cannot be the source of e/m radiation.

Let us specify to the hydrogen atom. The allowed Bohr orbits, on each of which the electron has a well-defined energy, correspond to the stationary states of quantum mechanics. In these states the electronic motion assumed by Bohr is equivalent to a constant electric current. Hence, in the "Bohr states" the atom does not radiate unless the electron makes a transition from a state of higher energy to a state of lower energy, in which case the atom will emit a photon of frequency proportional to the difference in energy between the two states.

---

[1] The advanced student may look into Chap. 3 of [4] (an old but classic textbook).

Now, when the hydrogen atom is not subject to external excitation, its electron "prefers" to be in the state of lowest energy, corresponding to the first Bohr orbit (the one closest to the nucleus). And, since no further transitions to states of lower energies are possible, the electron remains in the ground state and the atom no longer radiates. The energy of the atom stays fixed and the system avoids a catastrophic collapse. Stability is thus guaranteed.

So, by associating the semi-classical Bohr orbits with the stationary states of quantum mechanics, and by considering the quantum probability current as equivalent to an actual electric current, we are able to reconcile Bohr's theory with quantum mechanics and to explain, in essentially classical terms, why Bohr's atom is a stable system. The underlying idea is simple:

Stationary state ⇔ stationary current ⇔ no radiation ⇔ stability.

Even Maxwell wouldn't disagree!

## *6. In summary...*

In the initial stages of its development, atomic theory found itself in an awkward position trying to explain the stability of the atomic structure. The reason was that the modern picture of the atom seemed to violate the laws of classical electromagnetism, which dictate that every accelerating electric charge (here, the electron) must emit e/m radiation. And, if that were to happen in reality, the atom should collapse in almost no time! Atoms, however, are known to be stable structures.

The issue of stability was finally resolved by quantum mechanics, but at a price. Standard classical concepts such as the well-defined orbit of a particle had to be abandoned due to the uncertainty principle. Or, let us better say they had to be *reinterpreted*. Thus, the semi-classical orbits proposed by Bohr for the hydrogen atom were viewed as stationary states in the context of quantum mechanics.

And here comes a miracle: As a result of this conceptual redefinition, atomic stability may be "explained" in essentially classical terms in a way that is much more accessible to the non-specialist, compared to a full-blown quantum mechanical treatment of the problem. In simple words, stationary states are equivalent to time-independent currents. And, according to Maxwell's theory, such currents are not sources of radiation. The atom will rest comfortably in the ground state and will not collapse for lack of energy.

So, even in its classical (non-quantized) form, Maxwell's electromagnetism is essential for understanding the "logic" of quantum systems. Given that this theory also plays a fundamental role in relativity, one may justly regard J. C. Maxwell as probably the greatest theoretical physicist before Einstein! Notwithstanding the undisputed genius of Newton, I may add...

**References**

1.  C. J. Papachristou, *Introduction to Electromagnetic Theory and the Physics of Conducting Solids* (Springer, 2020).[2]

2.  C. J. Papachristou, *Electromagnetic waves, gravitational waves and the prophets who predicted them*, https://arxiv.org/abs/1603.00871

3.  C. J. Papachristou, *Introduction to Mechanics of Particles and Systems* (Springer, 2020).[3]

4.  J. J. Sakurai, *Advanced Quantum Mechanics* (Addison-Wesley, 1967).

---

[2] Manuscript: https://arxiv.org/abs/1711.09969

[3] Manuscript: http://metapublishing.org/index.php/MP/catalog/book/68